\pdfoutput=1
\documentclass{cpbtex}
\usepackage{graphicx}
\usepackage{dcolumn}
\usepackage{bm}
\usepackage{epstopdf}
\usepackage{ulem,xcolor}

\DeclareMathAlphabet{\mathsfsl}{OT1}{cmss}{m}{sl}

\newcommand{\me}{\mathrm{e}}
\newcommand{\mi}{\mathrm{i}}
\newcommand{\dif}{\mathrm{d}}

\begin{document}
\begin{CJK*}{GBK}{song}

\title{Achieving acoustic cloak by using compressible background flow\thanks{Project supported by the National Science Foundation of China (Grant No.11475088 and 11275024) and by the Ministry of Science and Technology of China (2013YQ030595-3).}}

\author{Ruo-Yang Zhang$^{1}$, \ Qing Zhao$^{2}$, \ and \ Mo-Lin Ge$^{1,2}$
\thanks{Corresponding author. E-mail:~geml@nankai.edu.cn}\\
$^{1}$Theoretical Physics Division, Chern Institute of Mathematics,\\ Nankai University, Tianjin,  300071, China\\
$^{2}${School of Physics, Beijing Institute of Technology, Beijing, 100081, China}}

\date{\today}
\maketitle

\begin{abstract}
  We propose a scheme of acoustic spherical cloaking by means of background irrotational flow in compressible fluid. The background flow forms a virtual curved spacetime and guides the sound waves bypass the cloaked objects. To satisfy the laws of real fluid, we show that spatially distributed mass source and momentum source are necessary to supply. The propagation of sound waves in this system is studied via both geometric acoustics approximation and full wave approach. The analytic solution of sound fields is obtained for plane wave incidence. The results reveal the effect of phase retardation (or lead) in comparison with the ordinary transformation-acoustic cloak. In addition, the ability of cloaking is also evaluated for unideal background flows by analyzing the scattering cross section.
\end{abstract}



\section{\label{sec:level1}Introduction}

Since the establishment of transformation optics~\cite{Pendry2006Sci,Leonhardt2006Sci,Schurig2006OE,Chen2010NatMat,pendry2015transforming,Leonhardt2010Dover}, the methodology for engineering wave functional devices via coordinate transformation has spread to various wave systems due to their common character, {\it i.e.} the form invariance of wave equations in arbitrary curvilinear coordinates and curved spacetimes. Transformation acoustics (TA) is one of these developments, which offers a standard procedure to design the material parameters so that sound waves propagate along prospective trajectories or hold proleptic properties in these materials~\cite{cummer2008material,chen2010acoustic}. The most representative device of TA is acoustic cloak that guides sound waves traveling around the cloaked objects without scattering and thus make itself and the hidden objects invisible from acoustic detections~\cite{cummer2007one,chen2007acoustic,cummer2008scattering,Norris2411Acoustic,Hua2009Hua}. From the standard approach of TA, acoustic cloaks in fluid are made up of anisotropic materials with tensor mass densities. Though real fluids do not possess these properties, a variety of cloaks have been produced by using artificial acoustic metamaterials~\cite{farhat2008broadband,Zhang2011broadband,popa2011experimental,sanchis2013three,zigoneanu2014three}.

On the other hand, it has been realized long ago that moving media act as curved spacetimes and can be used to simulate cosmic phenomena for both electromagnetic waves~\cite{gordon1923lichtfortpflanzung,leonhardt1999optics,leonhardt2000relativistic,de2003optical} and sound waves~\cite{white1973acoustic,unruh1981experimental,unruh1995sonic,visser1998acoustic,fischer2002riemannian,schutzhold2002gravity,unruh2008dumb,Bergliaffa2004Wave}.
In the pioneering works of Unruh~\cite{unruh1981experimental,unruh1995sonic}, he discovered the dynamic analogy between the massless particles in the spacetime of Schwarzschild black hole and the sound waves dominated by the radially flowing fluid. This analogy is also valid for superfluid~\cite{garay2000sonic,garay2001sonic,barcelo2001analogue,leonhardt2003bogoliubov} in which the effects associated with acoustic black hole have been observed experimentally~\cite{lahav2010realization,horstmann2010hawking,nguyen2015acoustic}. In recent years, Garc{\'\i}a-Meca {\it et~al.} combined the ideas of moving media and TA, and purposed the method of analogue transformation acoustics (ATA)~\cite{garcia2013analogue,garcia2014space,garcia2014analogue,garcia2014transformational}. They made use of background flow, instead of anisotropic effective media, to mimic the metric of spacetime, and to fabricate the desired acoustic devices, such as spacetime cloak and time-dependent spatial compressor.  Through forcing the effective metric to have a certain form, the parameters of background flow, {\it i.e.} the flow velocity, the speed of sound and the mass density, can be determined. However, Garc{\'\i}a-Meca {\it et~al.} never discussed the feasibility of these requisite parameters of background flow, although they are not unrestricted but submitted to the laws of fluid theoretically.

In this paper, we try to reconcile the ATA-determined parameters with the basic laws of fluid dynamics and show several constrains of the background flow.
First, the governing equation of sound waves in ATA is  accurate only when the background flow is irrotational, or at least the vorticity of the flow $\bm{\Omega}$ should be  much smaller than the frequency of sound $\omega$, in order that the analogy between the flow fields and curved spacetime is applicable~\cite{Bergliaffa2004Wave}.
Second, there should exist external force fields (momentum sources) and external mass sources in general to guarantee momentum and mass conservations of the background flow. Third, since the spatial distribution of pressure and mass density are fixed after ATA procedures, their relation restricts the state equation of the background fluid. As an example, we put forward a scheme for constructing spherical cloaks with background flow which complies with the realistic fluid dynamics. In addition, we investigate the propagation of sound in this cloaking system with both geometric acoustics approach and solving the wave equation analytically. Our results show that the major difference of our scheme from the standard TA cloak is the phase retardation (or lead) effect of the sound waves traveling around the cloak.  For eliminating the singularities of the parameters at the cloak surface, we introduce the unideal cloaks and evaluate their cloaking abilities.

\section{Basic theory}

To begin with, we briefly review the framework of ATA. The motion of ideal fluid (with no viscosity and thermal conductivity) obeys Euler equation and the continuity equation which correspond to the momentum and mass conservations respectively. If the fluid is barotropic, \textit{i.e.} its density $\rho$ only depends on pressure $P$, and the flow velocity $\bm{v}$ is irrotational, the compressible fluid submits to the following equations~\cite{LANDAU1987192}
\begin{subequations}\label{fluid}
\begin{flalign}
  &\frac{\partial \psi}{\partial t}+\frac{1}{2}\bm{v}^2+H(\rho)+U(\bm{r})=F(t),\\
  &\frac{\partial \rho}{\partial t}+\nabla\cdot\left(\rho\bm{v}\right)=M(\bm{r},t),
\end{flalign}
\end{subequations}
where $\psi$ is the velocity potential ($\bm{v}=\nabla\psi$), $H(\rho)=\int \dif P/\rho=\int \dif\rho\,c^2/\rho$ is the enthalpy per unit mass, $c=\sqrt{\dif P/\dif\rho}$ is the speed of sound, $F(t)$ is an arbitrary integral constant with respect to time, and we have supposed the existence of the external potential of momentum source $U(\bm{r})$ and the external mass source $M(\bm{r},t)$. Indeed, there are two origins of $U(r)$, one is the external bulk force $\bm{f}=-\nabla U_0$, the other corresponds to the mass source, thus $U=U_0+\int  M\bm{v}\cdot \dif\bm{r}$ (suppose $M\bm{v}$ is integrable).  As we will see afterward, the two external sources are necessary.

Considering sound waves traveling on a background flow with the parameters $\{\bm{u}, \psi, \rho,P\}$ obeying Eq.~(\ref{fluid}), the sound waves perform as the perturbations $\{\tilde{\bm{v}}, \tilde\psi, \tilde\rho,\tilde{P}\}$ of the flow. Taking the substitutions $\bm{v}\rightarrow\bm{u}+\tilde{\bm{v}}$, $\psi\rightarrow\psi+\tilde{\psi}$, $\rho\rightarrow\rho+\tilde{\rho}$ into Eq.~(\ref{fluid}), we can obtain the governing equations of sound wave in linear approximation~\cite{unruh1981experimental,unruh1995sonic,visser1998acoustic}
\begin{equation}\label{wave eq}
\begin{split}
   \partial_t\left(\frac{\rho}{c^2}\,\partial_t\tilde{\psi}\right)+\partial_t\left(\frac{\rho}{c^2}\,\bm{u}\cdot\nabla\tilde{\psi}\right)
   +\nabla\cdot\left(\frac{\rho}{c^2}\,\bm{u}\,\partial_t\tilde{\psi}\right)&\\
   +\nabla\cdot\left[\frac{\rho}{c^2}\,\bm{u}\left(\bm{u}\cdot\nabla\tilde{\psi}\right)\right]
   -\nabla\cdot\left(\rho\nabla\tilde{\psi}\right)=&0.
\end{split}
\end{equation}
And the sound pressure satisfies
\begin{equation}
\tilde{P}=c^2\tilde\rho=-\rho\left(\partial_t\tilde{\psi}+\bm{u}\cdot\nabla\tilde\psi\right).
\end{equation}
Eq.~(\ref{wave eq}) is identical with the D'Alembert equation in a curved spacetime
\begin{equation}\label{D'Alembert}
   \Box\,\tilde{\psi}=\frac{1}{\sqrt{-g}}\partial_\mu\left(\sqrt{-g}g^{\mu\nu}\partial_\nu\tilde{\psi}\right)=0,
\end{equation}
with the effective metric
\begin{equation}\label{}
   \left(g_{\mu\nu}\right)=\frac{\rho_1}{c_1}
   \left(
   \renewcommand\arraystretch{1.6}
   \begin{array}{c|@{\hspace{18pt}}c@{\hspace{18pt}}}
   -(c^2-u^2)  &    -u_{i}  \\ \hline
      -u_i     &  \gamma_{\,ij}
   \end{array}
   \right),
\end{equation}
where $\rho_1=\rho/\rho_0$, $c_1=c/c_0$, $\rho_0$ and $c_0$ are constants with dimension of density and velocity respectively,  $\gamma_{ij}$ denotes the spatial metric given in an arbitrary coordinate system of flat space, and $\dif x^0=\dif t$. This form of metric can be regarded as the generalized Schwarzschild metric written in the  Painlev\'{e}-Gullstrand type coordinate system (P-G system)~\cite{visser1998acoustic}. If $\bm{u}/(c^2-u^2)$ is integrable, we can obtain the  Schwarzschild type coordinate system (Sch.~system), through the coordinate transformation
\begin{equation}\label{transformation}
\dif\tau=\dif t+\frac{u_i\dif x^i}{c^2-u^2},\quad \dif x'^i=\dif x^i,
\end{equation}
where $\tau$ is the $x^0$ coordinate of Sch.~system. Then the metric turns into the standard form
\begin{equation}\label{metric in sch}
   \left(g'_{\mu\nu}\right)=\frac{\rho_1}{c_1}
   \left(
   \renewcommand\arraystretch{1.6}
   \begin{array}{c|c}
   -(c^2-u^2)        &   \mathbf{0}  \\\hline
       \mathbf{0}    &    \gamma_{\,ij}+\frac{u_iu_j}{c^2-u^2}
   \end{array}
   \right).
\end{equation}
In Sch.~system, time dimension and space dimensions are decoupled, nevertheless, the spatial projection of geodesics in this system holds the same form as in P-G system, since the spatial coordinates do not change. As pointed in Ref.~\cite{garcia2014transformational}, a major distinction between ATA and ordinary TA is that one generally deals with the wave equation of sound pressure in TA, while the kernel equation of ATA used to draw the analogy with curved spacetime is the equation of velocity potential. As a result, TA is restricted to design isobaric systems, whereas ATA is demanded to be globally barotropic~\cite{garcia2014transformational}.

According to the procedure of ATA, if we let the effective metric either in P-G system or in Sch. system equal to the requisite form gained by TA, the expressions of $\bm{u}$, $\rho$, $c$, can be determined. However, these quantities should also compose a real solution of fluid dynamics. By plugging this quantities back into Eq.~(\ref{fluid}), the required external sources are fixed. Besides, it is notable that the governing Eq.~(\ref{D'Alembert}) of ATA and Eq.~(\ref{metric in sch}) (if used) are valid only if $\bm{u}$ and $\bm{u}/(c^2-u^2)$ are irrotatinal. All these matters restrict the selective freedom of the background flow.

\begin{figure}[t]
\includegraphics[width=\columnwidth,clip]{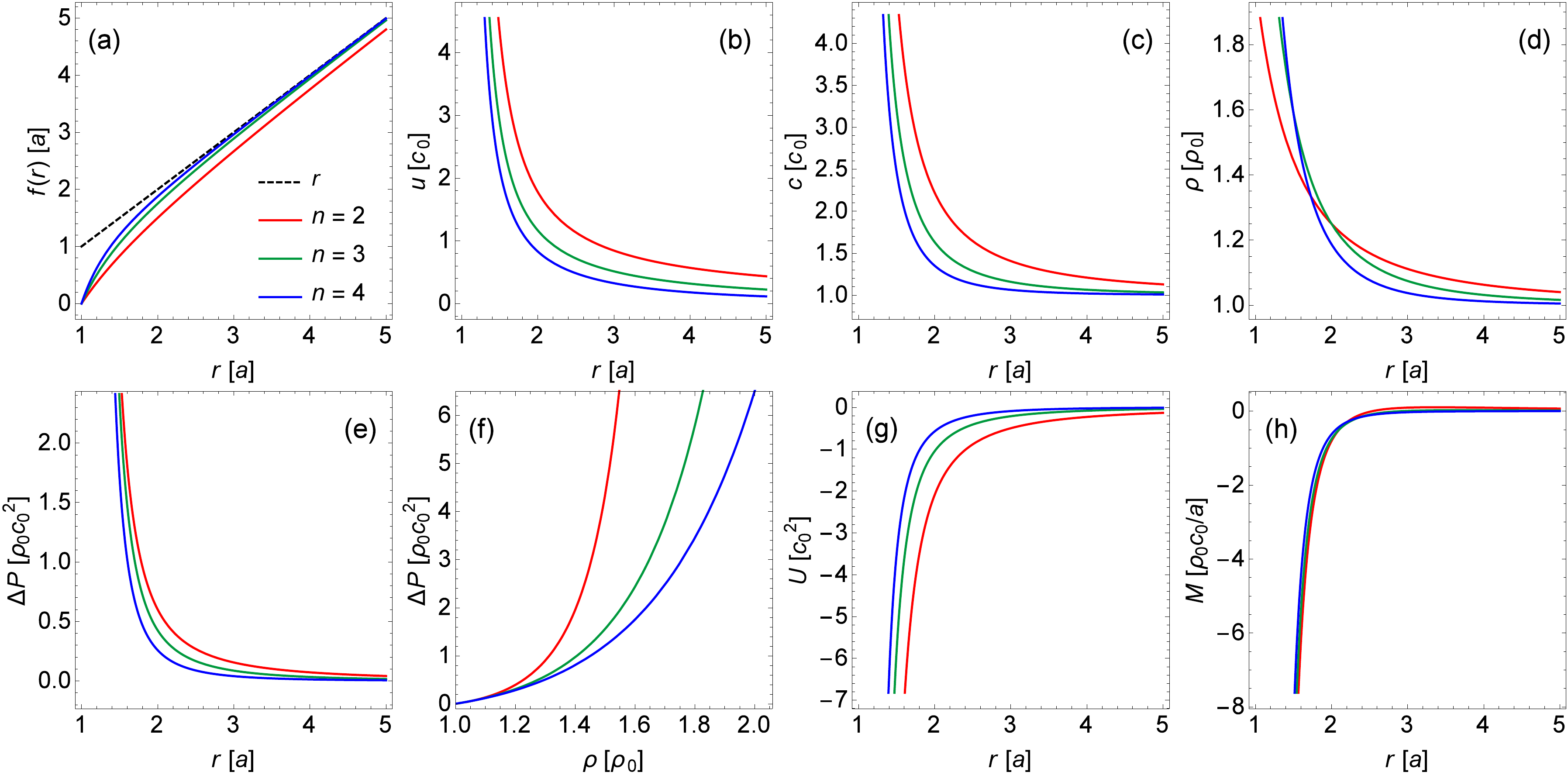}
\caption{\label{fig1}The profiles of (a) radial transformation $f(r)$, (b) velocity of background flow $u(r)$, (c) mass density $\rho$, (d) speed of sound $c(r)$, (e) pressure difference $\Delta P(r)=P-P_0$ in comparison with the pressure $P_0$ at infinite, (f) constitutive relation between $\rho$ and $\Delta P$, (g) external potential $U(r)$, (h) external mass source $M(r)$, where the red, green and blue curves correspond to the transformations given in Eq.~(\ref{f1}) with $n=2,\ n=3,\ \text{and}\ n=4 $ respectively. The quantity written in the square bracket beside each axis gives the unit of the corresponding coordinate axis. }
\end{figure}

Now we show how to construct a spherical cloak with background flow. For spherically symmetric systems, all parameters are merely functions of radius and the flow velocity $\bm{u}$ is radial. Consequently, $\bm{u}$ and $\bm{u}/(c^2-u^2)$ would always be integrable. It is convenient to discuss the problem in spherical coordinates $\{r,\theta,\phi\}$, thus we have $\left(\gamma_{ij}\right)=\mathrm{diag}(1,r^2,r^2\sin^2\theta)$ and $(u_i)=(u,0,0)$. In the light of TA, the effective metric for a spherical cloak reads~\cite{Leonhardt2010Dover}
 \begin{equation}
\left(g'_{\mu\nu}\right)=\textrm{diag}\left(-c_0^2,\ f'(r)^2,\ f(r)^2,\ f(r)^2\sin^2\theta\right).
\end{equation}
Here $f(r)$ ia an arbitrary radial transformation satisfying the invisibility condition $f(a)=0$, where $a$ is the inner radius of the cloak. And we demand $f(r)$ obeys the limit
$
\lim_{r\rightarrow \infty}\big(f(r)-r\big)=0,
$
so that the background flow goes to a homogenous and hydrostatic state with $\rho\rightarrow \rho_0,\ u\rightarrow0,\ c\rightarrow c_0,\ P\rightarrow P_0$, when $r\rightarrow\infty$. Letting Eq.~(\ref{metric in sch}) equal to the effective metric of spherical cloak, we obtain
\begin{subequations}\label{relations}
\begin{align}
  &\rho\ =\ \rho_0f'(r),\label{rho}\\
  &c\ =\ c_0\frac{r^2f'(r)}{f(r)^2},\\
  &\bm{u}\  =\ \pm\,c_0\frac{r}{f(r)^2}\sqrt{r^2f'(r)^2-f(r)^2}\ \bm{\hat{e}}_r.\label{u}
\end{align}
\end{subequations}
The signs $\pm$ of $\bm{u}$ indicate that the flow could be either emanative or convergent.  In terms of the barotropic postulate, the pressure can be derived by
\begin{equation}\label{pressure}
P(r)=\int c^2\dif\rho=c_0^2\rho_0\int_{\infty}^r \frac{r^4f'(r)^2f''(r)}{f(r)^4}\dif r+P_0.
\end{equation}
We can further obtain the required state equation of the fluid, namely the constitutive relation between $P$ and $\rho$, in terms of Eq.~(\ref{rho}) and Eq.~(\ref{pressure}). Substituting  Eq.~(\ref{relations}) into Eq.~(\ref{fluid}), we get the required external potential and mass source
\begin{subequations}
\begin{align}
  \begin{split}
  &U(r) = -\frac{1}{2}\bm{u}^2-H(\rho)\\
       = &-\frac{c_0^2r^2}{2f(r)^4}\left(r^2f'(r)^2-f(r)^2\right)-c_0^2\int \frac{r^4f'(r)f''(r)}{f(r)^4}\dif r,
  \end{split}\\
    \begin{split}
  &M(r) = \nabla\cdot(\rho\bm{u})\\
       = &\pm\frac{\rho_0c_0}{r^2}\frac{\dif}{\dif r}\left(\frac{r^3f'(r)}{f(r)^2}\sqrt{r^2f'(r)^2-f(r)^2}\right).
  \end{split}
\end{align}
\end{subequations}
Obviously, the desired background flow would not generally satisfy the momentum and mass conservations unless the above external source terms are provided. In principle, the external potential could be made up with external electric fields acting on charged fluid. The mass sources could be realized via the mass exchange of chemical reactions or via phase transition of multiphase fluid~\cite{asfaw2010prandtl}, or even via discretely distributed jets and outlets.

A class of functions satisfying the conditions $f(a)=0$ and $f(r)\rightarrow r$ as $r\rightarrow\infty$ is
\begin{equation}\label{f1}
   f(r)=r-\frac{a^n}{r^{n-1}},\qquad (n\geq2).
\end{equation}
We calculate the profiles of $u,\ \rho,\ c,\ P,\ M$ with respect to $r$ for $n=2,3,4$, and obtain the required state equation $P(\rho)$. The results are shown in Fig.~\ref{fig1}. The figures illustrate the case of outgoing flow from the center ($\bm{u}$ holds the sign of $``+''$ in Eq.~(\ref{u})). In this case, the external potential is attractive and the mass source is negative. If $\bm{u}$ is convergent to the center, the mass source should be positive while other variables do not change.

\section{Geometric acoustics}

In this section, we study the propagation  of sound waves with geometric acoustics approximation in the shortwave limit. Under the eikonal hypothesis, the velocity potential takes the form  $\tilde{\psi}\sim \me^{\mi\Phi(x^\mu)}$, and the phase $\Phi$ satisfies the eikonal equation $H=h(x^\mu)g^{\mu\nu}k_\mu k_\nu=0$, where $k_\mu=\partial_\mu \Phi$, $h(x^\mu)$ can be an arbitrary non-degenerate function, and
\begin{equation}\label{}
\begin{split}
   \left(g^{\mu\nu}\right)&=\frac{1}{c_0c\rho_1}
   \left(
   \renewcommand\arraystretch{1.6}
   \begin{array}{@{\hspace{18pt}}c@{\hspace{18pt}}|c}
   -1  &    -u^{i}  \\ \hline
      -u^i     &  c^2\gamma^{ij}-u^iu^j
   \end{array}
   \right)\\
   &=\frac{1}{c_0c\rho_1}\ \left(\tilde{g}^{\mu\nu}\right).
\end{split}
\end{equation}
Sound rays obey the following canonical equations
\begin{equation}\label{canonical equations}
\frac{\dif x^\mu}{\dif s}=\frac{\partial H}{\partial k_\mu}, \qquad \frac{\dif k_\mu}{\dif s}=-\frac{\partial H}{\partial x^\mu}.
\end{equation}
 It can be verified that the formulas of sound rays are identical with the geodesics in the P-G system with the effective metric. If $g^{\mu\nu}$ is time-independent, $\dif k_0/\dif s=-\partial H/\partial t\equiv0$, thus the frequency $\omega=-k_0$ is conserved along a ray.
Letting $h(x^\mu)=c_0c\rho_1$, $H=\tilde{g}^{\mu\nu}k_\mu k_\nu=c^2k^2-(\omega-\bm{u}\cdot\bm{k})^2=0$, we obtain  $\omega=\bm{u}\cdot\bm{k}+ck$,
which is exactly the dispersion relation of sound waves in moving media~\cite{LANDAU1987192}, and is always applicable in the limit of geometric acoustics no matter if the background flow are irrotational or not. Therefore, although the analogy between background flow and the curved spacetime is rigorous only for potential flow, nevertheless, this restriction can be drawn off in the limit of geometric acoustics. In fact, as proved in Ref.~\cite{Bergliaffa2004Wave}, the analogy is still valid as long as the frequency of the sound is larger enough than the vorticity of the flow $\bm{\Omega}=\nabla\times\bm{u}$: $\omega\gg|
\bm{\Omega}|$, but there is no requirement of the magnitude of the spatial inhomogeneity  comparing with wave length.

Substituting the Hamiltonian into the first set of the canonical equations, we obtain
\[
    \frac{\dif t}{\dif s} =2kc,\qquad \frac{\dif x^i}{\dif s} =2kc\left(u^i+ck^i/k\right).
\]
\begin{figure}
\begin{center}
\includegraphics[width=0.6\columnwidth,clip]{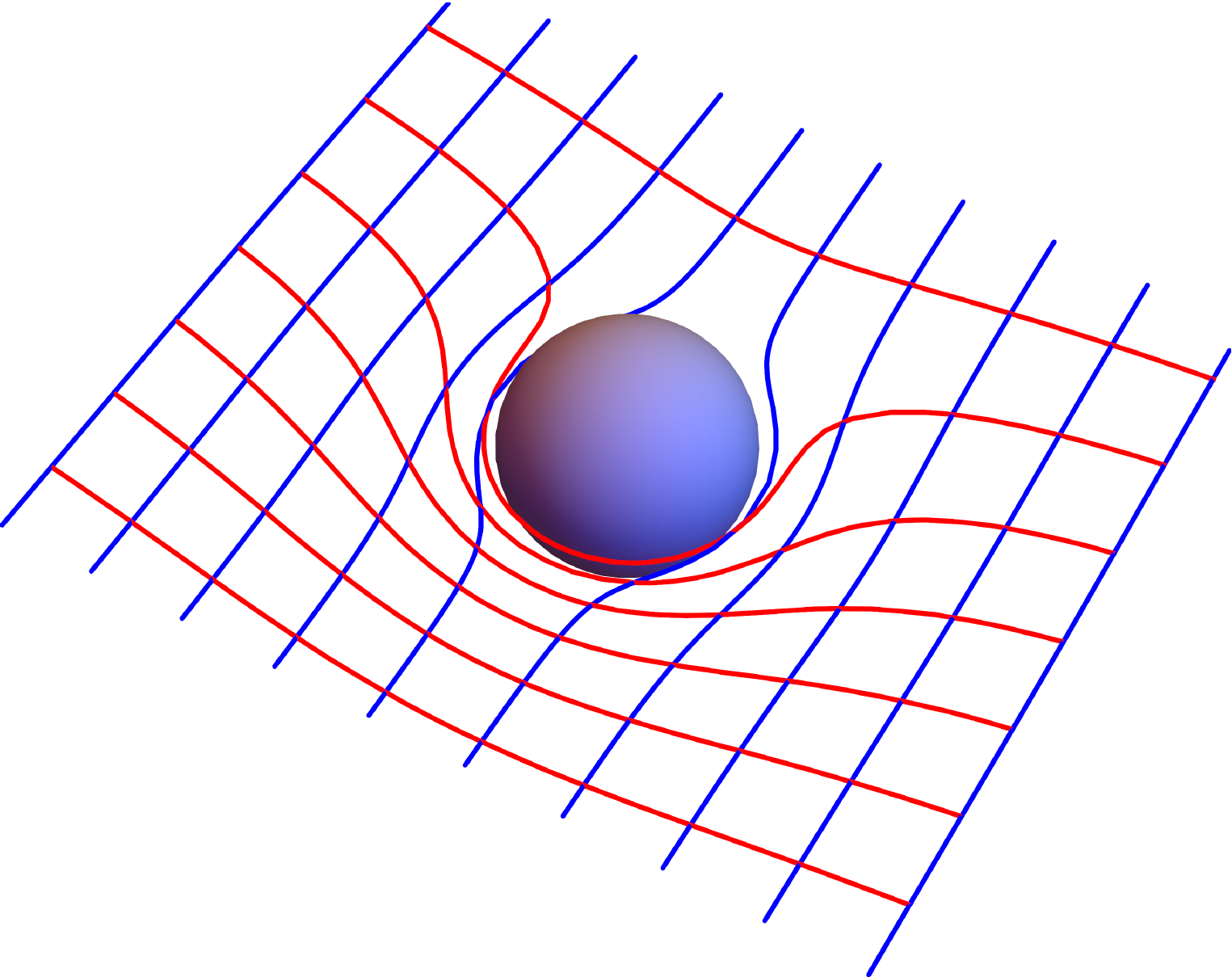}
\end{center}
\caption{\label{fig2}Sound rays in the acoustic cloaking system. The blue curves denote the sound rays, and the red curves denote wave fronts, where the background flow (the direction is emanative from the center) corresponds to Eq.~(\ref{f1}) with $n=4$. }
\end{figure}
Cancellation of $ds$ leads to
\begin{equation}\label{}
\frac{\dif x^i}{\dif t} =u^i+ ck^i/k.
\end{equation}
The result is identical with the group velocity of the sound wave, $\bm{v}_g=\nabla_{\bm{k}}\omega=\bm{u}+ c\bm{k}/k$. Actually, for arbitrary eikonal equations $H(-\omega,k_i,x^\mu)=0$,
\[
v_g^i= \frac{\partial \omega}{\partial k_i}=-\frac{\partial H/\partial k_i}{\partial H/\partial \omega}=\frac{\partial H/\partial k_i}{\partial H/\partial k_0}.
\]
Therefore, the group velocity is always along the solution of the canonical equations.

For spherically symmetric metric, $k_\phi$ is conserved along a ray, since $dk_\phi/ds=0$. If we let $k_\phi=0$, then $\phi\equiv\phi_0,$ and $dk_\theta/ds=0$. As a result, the angular momentum of phonon perpendicular to the orbit $\ k_\theta\equiv L$ is also conserved. We can further remove all the wave vector terms in the canonical equations, and obtain
\begin{subequations}
\begin{gather}
  \left(1-\frac{u^2}{c^2}\right)\frac{\dif t}{\dif\sigma}+\frac{u}{c^2}\frac{\dif r}{\dif\sigma}=\omega,\label{t}\\
  \left(\frac{\dif r}{\dif\sigma}\right)^2=\frac{\omega^2}{c^2}- \left(1-\frac{u^2}{c^2}\right)\frac{L^2}{r^2},\label{r}\\
  \qquad\frac{\dif\theta}{\dif\sigma}=\frac{L}{r^2},\quad\  \phi\equiv\phi_0,\label{theta}
\end{gather}
\end{subequations}
with $\dif\sigma=2c^2\dif s$. Inserting Eq.~(\ref{theta}) into Eq.~(\ref{r}), we get the compact form of the geodesic equation
\begin{equation}\label{ray equation}
\begin{split}
  \left(\frac{1}{r^2}\frac{\dif r}{\dif\theta}\right)^2 &=\frac{\omega^2}{L^2}\frac{1}{c^2}-\left(1-\frac{u^2}{c^2}\right)\frac{1}{r^2}.
\end{split}
\end{equation}
Substitution the parameters of spherical cloak yields
\begin{equation}
   \dif\theta=\pm \frac{\dif f}{f\sqrt{\frac{\omega^2}{L^2c_0^2}f^2-1}}.
\end{equation}
By integration, we obtain the general analytic solution of the sound rays
\begin{equation}\label{analytic ray function}
f(r)\sin(\theta-\theta_0)=\pm Lc_0/\omega,\quad \phi\equiv\phi_0,
\end{equation}
where $Lc_0/\omega$ denotes the impact parameter of the phonon incident from infinity. Actually, Eq.~(\ref{analytic ray function}) is exactly transformed from the straight  lines $r\sin(\theta-\theta_0)=\mathrm{const.}$ according to TA. Moreover, based on Eq.~(\ref{ray equation}), the sound rays only depend on $\bm{u}$ and $c$ but are independent of the density distribution $\rho$, therefore, we have a freedom to select $\rho$ in the limit of geometric acoustics.

Fig.~\ref{fig2} displays a set of rays (blue curves) incident with the same direction but different impact parameters. Despite the same trajectories as in TA cloak, the wave fronts $\Phi=\mathrm{const.}$, shown by the red curves in Fig.~\ref{fig2}, significantly differ from those in TA cloak. In our case, the wave fronts before bypassing the cloak are asymmetric to those after bypassing the cloak, however, this symmetry exists in TA cloak~\cite{Schurig2006OE,cummer2007one,chen2007acoustic,cummer2008scattering}. This effect presents the phase retardation of the waves when bypassing the cloak, and it is essentially induced by the background flow $\bm{u}$. In the next section, we will further investigate this effect.

\section{Analytic solution}

According to Eq.~(\ref{transformation}), by applying the rules
\[
\partial_t\rightarrow\partial_\tau,\quad \partial_i\rightarrow\partial_i+\frac{u_i}{c^2-u^2}\partial_\tau,
\]
the wave equation (\ref{wave eq}) in Sch. system takes the form
\begin{equation}\label{wave eq2}
   \frac{\rho}{c^2-u^2}\,\partial_\tau^2\,\tilde{\psi}-\frac{1}{\sqrt{\gamma}}\,\partial_i\left[\sqrt{\gamma}\,\frac{\rho}{c^2}\left(c^2\gamma^{ij}-u^iu^j\right)\partial_j\,\tilde{\psi}\right]=0.
\end{equation}
Substituting Eqs.~(\ref{relations}) into Eq.~(\ref{wave eq2}), we have
\begin{equation}\label{wave eq3}
\begin{split}
   \left\{\frac{1}{f^2}\frac{\partial}{\partial f}\left(f^2\frac{\partial \tilde{\psi}}{\partial f}\right)+\frac{1}{f^2\sin\theta}\frac{\partial}{\partial \theta}\left(\sin\theta\frac{\partial\tilde{\psi}}{\partial\theta}\right)\right.&\\
   \left.+\frac{1}{f^2\sin^2\theta}\frac{\partial^2\tilde{\psi}}{\partial\phi^2}\right\}-\frac{1}{c_0^2}\,\frac{\partial^2}{\partial\tau^2}\,\tilde{\psi}&=0.
\end{split}
\end{equation}
Considering monofrequent  wave $\tilde{\psi}(\bm{r},t)=\tilde{\psi}_0(\bm{r})\me^{-\mi\omega t}$, since $t=\tau-\int^r_b\frac{u\dif r}{c^2-u^2}$, the velocity potential can be expressed as
$
\tilde{\psi}(\bm{r},t)
=\tilde{\psi}(\bm{r})\me^{-\mi\omega \tau}
$
with $\tilde{\psi}(\bm{r})=\tilde{\psi}_0(\bm{r})\exp\left[\mi\,\omega\int^r_b\frac{u\dif r}{c^2-u^2}\right]$. Then Eq.~(\ref{wave eq3}) becomes
\begin{equation}\label{wave eq in sch}
\begin{split}
   \frac{1}{f^2}\frac{\partial}{\partial f}\left(f^2\frac{\partial \tilde{\psi}}{\partial f}\right)+\frac{1}{f^2\sin\theta}\frac{\partial}{\partial \theta}\left(\sin\theta\frac{\partial\tilde{\psi}}{\partial\theta}\right)&\\
   +\frac{1}{f^2\sin^2\theta}\frac{\partial^2\tilde{\psi}}{\partial\phi^2}+\tilde{k}_0^2\tilde{\psi}&=0,
\end{split}
\end{equation}
where $\tilde{k}_0=\omega/c_0$ and $\tilde{\psi}$ is the abbreviation of $\tilde{\psi}(\bm{r})$. By separation of variables, $\tilde{\psi}(\bm{r})=R(f)Y(\theta,\phi)$, the radial equation of $R(f)$ is spherical Bessel equation, and the angular equation is spherical harmonious equation. Therefore, the solutions can be expressed by
\begin{subequations}\label{solution}
\begin{flalign}
  \tilde{\psi}^{\mathrm{in}}\ &=\ \sum_{lm}\left[A^{\mathrm{in}}_l\, j_l(\tilde{k}_0f)+B^{\mathrm{in}}_l\, n_l(\tilde{k}_0f)\right]Y_{lm}(\theta,\phi),\\
  \tilde{\psi}^{\mathrm{s}}\ &=\ \sum_{lm}A^{\mathrm{s}}_l\, h^{(1)}_l(\tilde{k}_0f)Y_{lm}(\theta,\phi),
\end{flalign}
\end{subequations}
where $\tilde{\psi}^{\mathrm{in}}$ and $\tilde{\psi}^{\mathrm{s}}$ are the velocity potentials of incident wave and scattering wave respectively, $j_l$, $n_l$, $h^{(1)}_l$ denote the $l$ order spherical Bessel, Neumann, and first Hankel functions respectively,  and $Y_{lm}(\theta,\phi)$ denotes the $(l,m)$ order spherical harmonics. We demand that the incident wave tends to plane wave at infinity:
\begin{equation}\label{}
\begin{split}
   \lim_{r\rightarrow\infty}\tilde{\psi}^{\mathrm{in}}\me^{-\mi\omega\tau} & \sim Ae^{i(\tilde{k}_0r\cos\theta-\omega t)}\\
   & =\sum_{l=0}^\infty a_l j_l(\tilde{k}_0r)P_l(\cos\theta)\me^{-\mi\omega t},
\end{split}
\end{equation}
with $a_l=A(2l+1)\mi^l$. The incident wave thus can be written as
\begin{align}\label{boundary condition}
      &\tilde{\psi}^{\mathrm{in}}(\bm{r},t) = \sum_{l=0}^\infty \tilde{a}_l j_l(\tilde{k}_0f)P_l(\cos\theta)\me^{-\mi\omega\tau}\nonumber\\
    = &A\exp\left[\mi\tilde{k}_0f(r)\cos\theta-\mi\omega\int^r_\infty\frac{u\dif r}{c^2-u^2}\right]\me^{-\mi\omega t}\\ \
    = &A\exp\left[\mi\tilde{k}_0\left(f(r)\cos\theta\mp\int^r_\infty\frac{\sqrt{r^2f'^2-f^2}}{r}\dif r\right)\right]\me^{-\mi\omega t},\nonumber
\end{align}
where $\tilde{a}_l=a_l\exp\left[\int^\infty_b\frac{u\dif r}{c^2-u^2}\right]$,  $P_l(\cos\theta)$ is the $l$ order Legendre polynomial, and the signs $\mp$ in the last line correspond to  emanative and convergent flows respectively (the signs $\mp$ have the same meaning in all of the following equations). At the inner interface $r=a$, we suppose there is a hard boundary of the sound wave:
\begin{gather}
   \left.\left(v^{\mathrm{in}}_r+v^{\mathrm{s}}_r\right)\right|_{r=a}=\left.\left(\frac{\partial\tilde{\psi}^{\mathrm{in}}_0}{\partial r}+\frac{\partial\tilde{\psi}^{\mathrm{s}}_0}{\partial r}\right)\right|_{r=a}=0.\\
    \text{(P-G system)}\notag
\end{gather}
\begin{figure}[!t]
\centering
\includegraphics[width=0.9\columnwidth,clip]{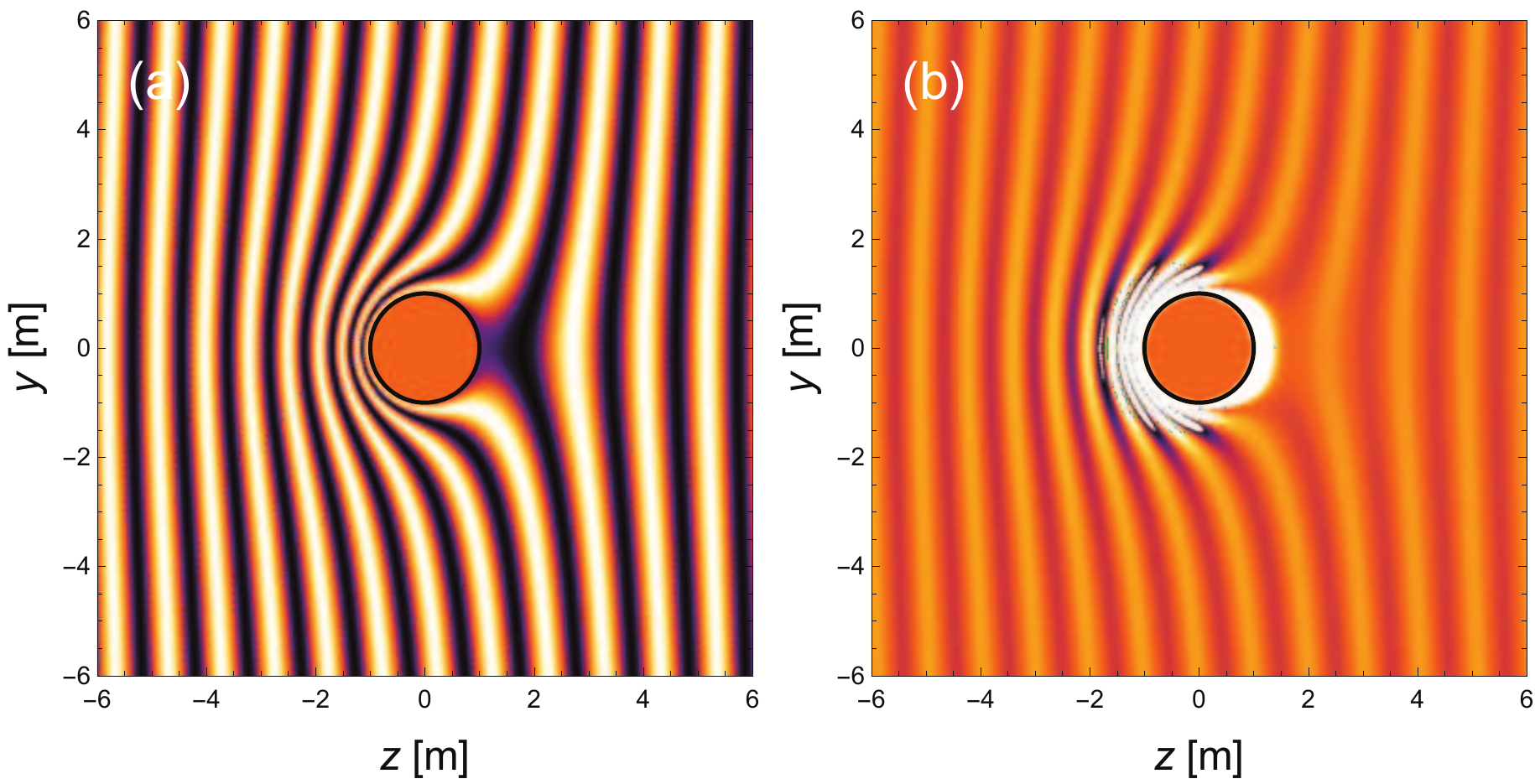}
\caption{\label{fig3}Distributions of (a) velocity potential $\tilde{\psi}$, (b) sound pressure $\tilde{P}$, in $x=0$ plane, of the sound wave incident from the left side, where the background flow (emanative from the center) corresponds to Eq.~(\ref{f1}) with $n=4$.}
\end{figure}
In Sch. system, the boundary condition turns into
\begin{gather}
   \left(\frac{\partial}{\partial r}+\frac{u}{c^2-u^2}\frac{\partial}{\partial\tau}\right)\left.\left(\tilde{\psi}^{\mathrm{in}}_r+\tilde{\psi}^{\mathrm{s}}_r\right)\right|_{r=a}=0.\\
   \text{(Sch. system)}\notag
\end{gather}
Then the scattering wave is derived as
\begin{equation}\label{scattering}
\begin{split}
   \tilde{\psi}^{\mathrm{s}}(\bm{r},t) = &\sum_{l=0}^\infty \frac{\tilde{j}'_l(x_a)}{\tilde{h}^{(1)\prime}_l(x_a)}\tilde{a}_l h^{(1)}_l(\tilde{k}_0f)P_l(\cos\theta)\me^{-\mi\omega\tau}\\
    = &\sum_{l=0}^\infty \frac{\tilde{j}'_l(x_a)}{\tilde{h}^{(1)\prime}_l(x_a)}a_l h^{(1)}_l(\tilde{k}_0f)P_l(\cos\theta)\\
    &\cdot\exp\left[\mp\,\mi\tilde{k}_0\int^r_\infty\frac{\sqrt{r^2f'^2-f^2}}{r}dr\right]\me^{-\mi\omega t},
\end{split}
\end{equation}
with $x_a=\tilde{k}_0f(a)$, and
\begin{subequations}
\begin{flalign}
  \tilde{j}'_l(x_a)&=j'_l(x_a)\mp i\textstyle{\sqrt{1-\frac{f(a)^2}{a^2f'(a)^2}}}j_l(x_a),\\
  \tilde{h}^{(1)\prime}_l(x_a)&=h^{(1)\prime}_l(x_a)\mp i\textstyle{\sqrt{1-\frac{f(a)^2}{a^2f'(a)^2}}}h^{(1)}_l(x_a).
\end{flalign}
\end{subequations}
If $f(a)=0$, the scattering wave $\tilde{\psi}^{\mathrm{s}}\equiv0$. Consequently, we accomplish ideal acoustic invisibility cloaking by means of background flow. The sound pressure can be obtained from
\begin{equation}\label{sound pressure}
   \tilde{P}(\bm{r},t) = \rho\left(i\omega-u\frac{\partial}{\partial r}\right)\left(\tilde{\psi}^{\mathrm{in}}(\bm{r},t)+\tilde{\psi}^{\mathrm{s}}(\bm{r},t)\right).
\end{equation}
For the ideal case, Eq.~(\ref{sound pressure}) reduces to
\begin{align}\label{sound pressure2}
    &\tilde{P}(\bm{r},t)\nonumber\\
   =  &\mi\omega\rho_0f'^2\left(f'\mp\frac{r}{f^2}\sqrt{r^2f'^2-f^2}\cos\theta\right)\\
   \cdot & A\exp\left[\mi\,\tilde{k}_0\left(f(r)\cos\theta\mp\int^r_\infty\frac{\sqrt{r^2f'^2-f^2}}{r}dr\right)\right]\me^{-\mi\omega t}.\nonumber
\end{align}
\begin{figure}[!t]
\centering
\includegraphics[width=0.9\columnwidth,clip]{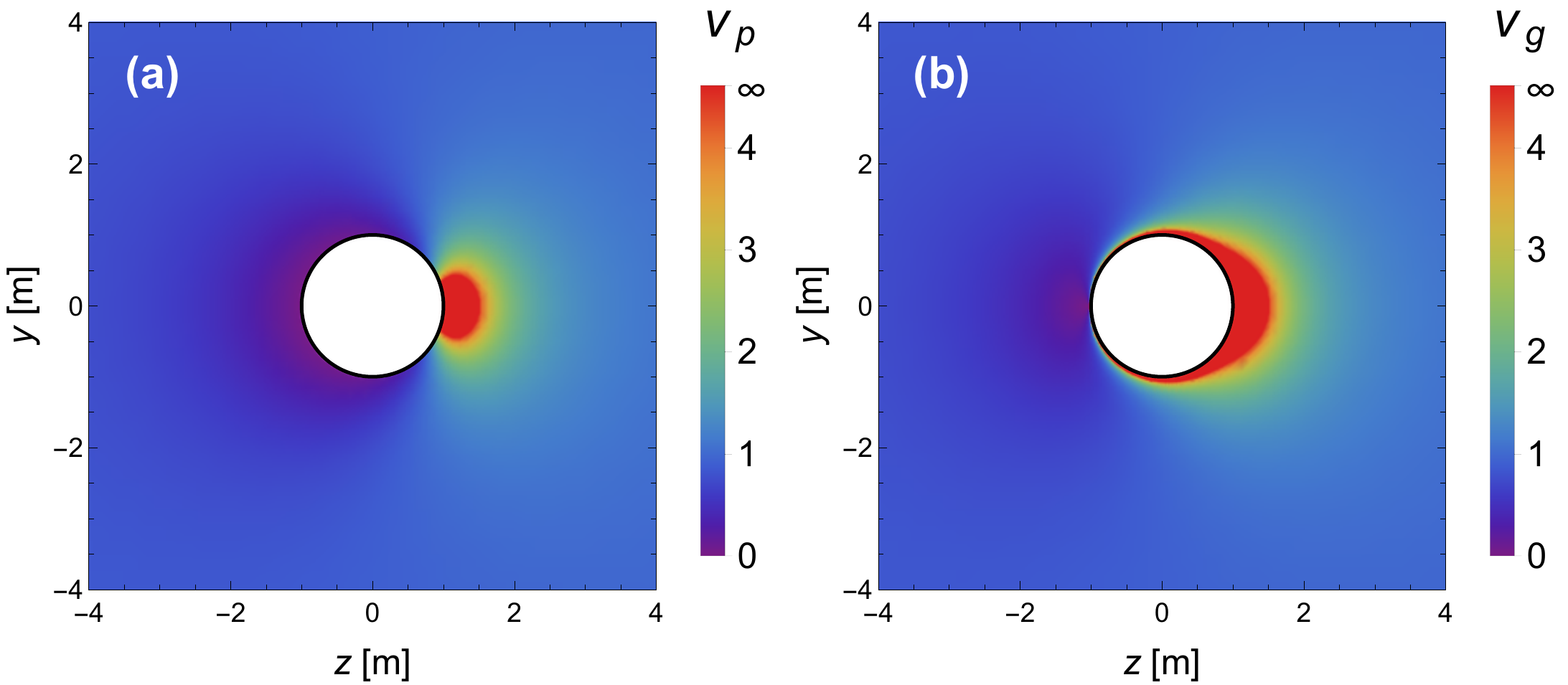}
\caption{\label{fig4}Distributions of (a) phase velocity $v_p$, (b) group velocity $v_g$, in $x=0$ plane, of the sound wave incident from the left side, where the background flow (emanative from the center) corresponds to Eq.~(\ref{f1}) with $n=4$.}
\end{figure}
Fig.~\ref{fig3} shows the field distributions of both velocity potential $\tilde\psi$ and sound pressure $\tilde{P}$ for the case of idea cloaking with emanative background flow. As shown in the figures, the fields are asymmetric with respect to the $z=0$ plane in contrast to the case of TA cloak. For the incident half place ($z<0$), the phases of the fields are compressed heavily, while the phases are stretched in the outgoing half place ($z>0$). According to the expression of the wave fronts
\begin{equation}\label{wave front}
   \Phi=\tilde{k}_0\left[f(r)\cos\theta\mp\int^r_\infty\frac{\sqrt{r^2f'^2-f^2}}{r}\dif r\right]=\text{const.},
\end{equation}
the phase retardation and lead originates exactly from the additional phase $\int^r_\infty\frac{u\dif r}{c^2-u^2}=\int^r_\infty\frac{\sqrt{r^2f'^2-f^2}}{r}\dif r$. Besides, if $u$ is convergent to the center, the effect turns into opposite, namely the phases are stretched in the incident half place while are compressed in the outgoing half, and the field distributions shown in Fig.~\ref{fig3} are identical with the situation of a wave incident form the right side.
This effect also can be viewed in the light of the sound velocities. According to Eq.~(\ref{sound pressure2}), the wave vector reads
\begin{equation}
\begin{split}
    &\bm{k}=\nabla\Phi\\
    =&\tilde{k}_0\left[\left(f'\cos\theta\mp\frac{1}{r}\sqrt{r^2f'^2-f^2}\right)\bm{\hat{e}}_r-\frac{f\sin\theta}{r}\,\bm{\hat{e}}_\theta\right].
\end{split}
\end{equation}
Hence, the phase velocity of the sound can be obtained by $v_p=\omega/k$ with $k=\tilde{k}_0\big(f'\mp\sqrt{r^2f'^2-f^2}\cos\theta/r\big)$. Note that $c$ is the phase velocity of sound wave in static fluid, but $v_p\neq c$ in moving media.
And the group velocity is
\begin{equation}
     \bm{v}_g=\bm{u}+ c\bm{k}/k=v_p\left(\cos\theta\bm{\hat{e}}_r-\frac{rf'(r)}{f(r)}\bm{\hat{e}}_\theta\right).
\end{equation}
Fig.~\ref{fig4} exhibits the distributions of phase velocity and group velocity corresponding to the sound fields given in Fig.~\ref{fig3}. As we can see, the phase velocity and the group velocity slow down as the sound wave approaches the inner boundary $r=a$ from the left side, whereas the velocities have extremely large values at the right side of inner boundary. For the case of convergent back ground flow, the distributions in Fig.~\ref{fig4} should yet correspond to the wave incident form the right side.  It is precisely the asymmetric distribution of phase velocity that causes the phase retardation and lead in corresponding regions.

Noteworthy, an alternative scheme of acoustic cloak based on velocity potential wave equation has been porposed in Ref.~\cite{garcia2014transformational}.
In spite of the remarkable differences between their results and the ordinary TA cloaks treated with sound pressure equation, their proposal is more similar to a standard TA device than ours, since their primary method is still to construct effective anisotropic media with the conventional procedure used in TA~\cite{Pendry2006Sci,Schurig2006OE}, but our key idea totally originates from the analogy between the moving media and curved spacetime.

\begin{figure}[!t]
\centering
\includegraphics[width=0.8\columnwidth,clip]{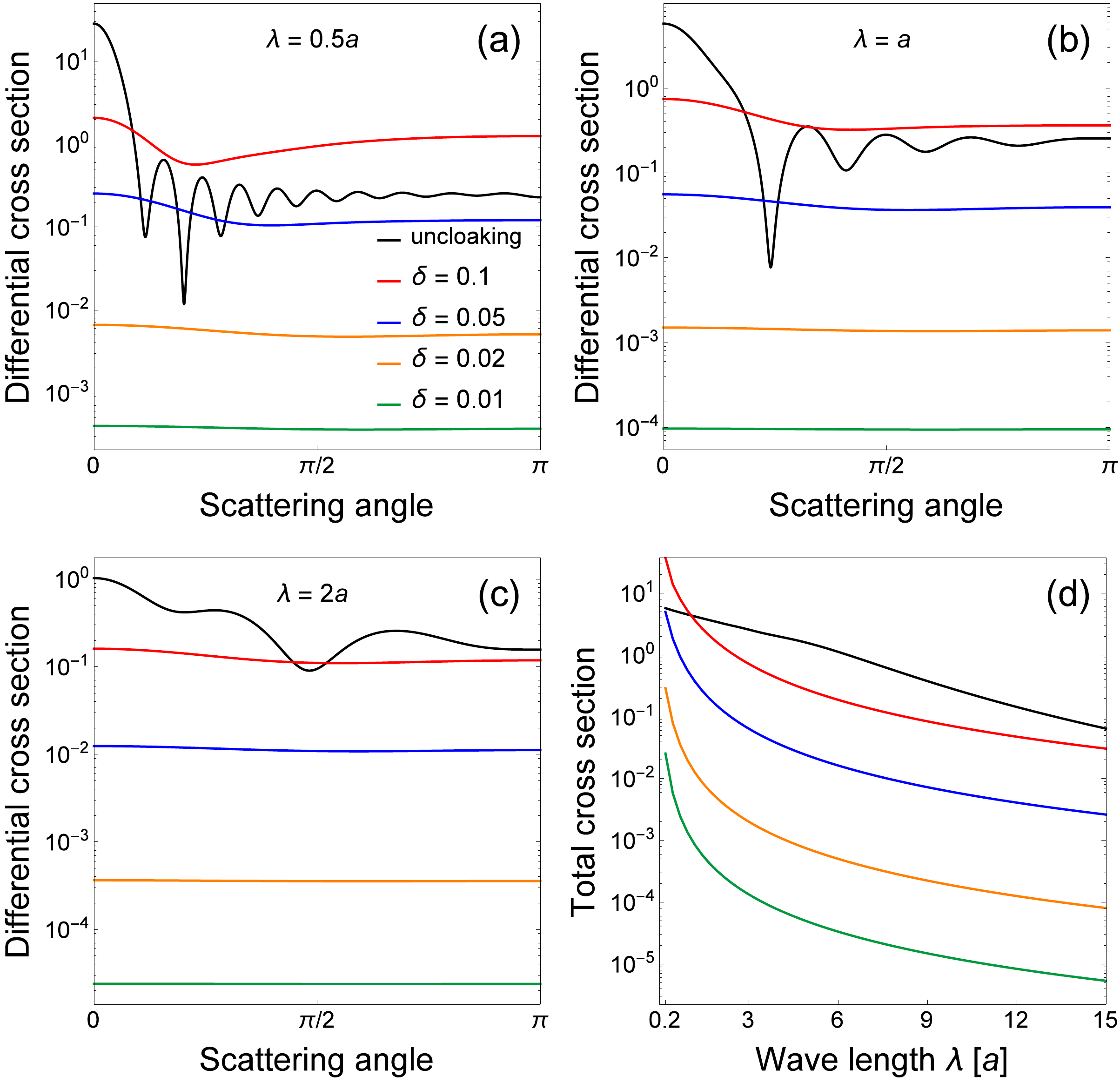}
\caption{\label{fig5}The angular distributions of differential scattering cross sections of unideal cloaks at the wave lengthes: (a) $\lambda=0.5a$, (b) $\lambda=a$, (c) $\lambda=2a$. (d) The total scattering cross section versus wave length. In each figure, the black curve denotes the scatter of a ball with unit radius and hard boundary, the red, blue, orange, green curves correspond to the unideal cloaks of $\delta=0.1,\,0.05,\,0.02,\,0.01$ respectively.  }
\end{figure}

\section{Unideal cloaking}

To realize ideal cloaking, all parameters of the background flow should tend to infinity at the inner boundary $r=a$ as shown in Fig.~\ref{fig1}, however, it is impossible in practice.  In this section, we would discuss the unideal case with simply cutting off the infinite parameters. We still use the radial transformation $f(r)$ to construct the background flow, but let its zero point be $a'=(1-\delta)a$ in order to escape the singularity at the inner boundary, when $\delta\rightarrow0$  the cloak approaches to an ideal one. In terms of Eq.~(\ref{scattering}), the scattering wave tends to
\begin{equation}
  \tilde{\psi}^{\rm s}\rightarrow A\frac{\me^{\mi\,kr}}{kr}\sum_{l=0}^\infty (2l+1)\mi\, A_l P_l(\cos\theta), \quad \text{as }r\rightarrow\infty,
\end{equation}
with $A_l=\tilde{j}'_l(x_a)/\tilde{h}_l^{(1)\prime}(x_a)$. Therefore, we can evaluate the cloaking ability of the background flow by means of the differential scattering cross section (DSCS)
\begin{equation}
  \frac{\dif\sigma}{\dif\Omega}=\left|\frac{1}{k}\sum_{l=0}^\infty(2l+1)\mi A_lP_l(\cos\theta)\right|^2,
\end{equation}
and the total scattering cross section (TSCS) $\sigma=\oint\frac{\dif\sigma}{\dif\Omega}\sin\theta{\dif\phi}{\dif\theta}$. We have checked the angular distributions of DSCS, and TSCS changing with wave length for $\delta=0.1$, $\delta=0.05$, $\delta=0.02$, and $\delta=0.01$ in comparison with the uncloaking results, as shown in Fig.~\ref{fig5}. The results reveal that the unideal constructions still have good property of cloaking for nearly full band of wave length as $\delta\leq 0.05$. However, for the case of $\delta=0.1$, only when $\lambda>2a$, its invisibility effect  is acceptable. In addition, the angular distributions of DSCS show that the wave scatters quite uniformly in all directions when the background flow exists, whereas there are usually  large components of back scattering in the uncloaking cases.

\begin{figure}[!t]
\centering
\includegraphics[width=0.6\columnwidth,clip]{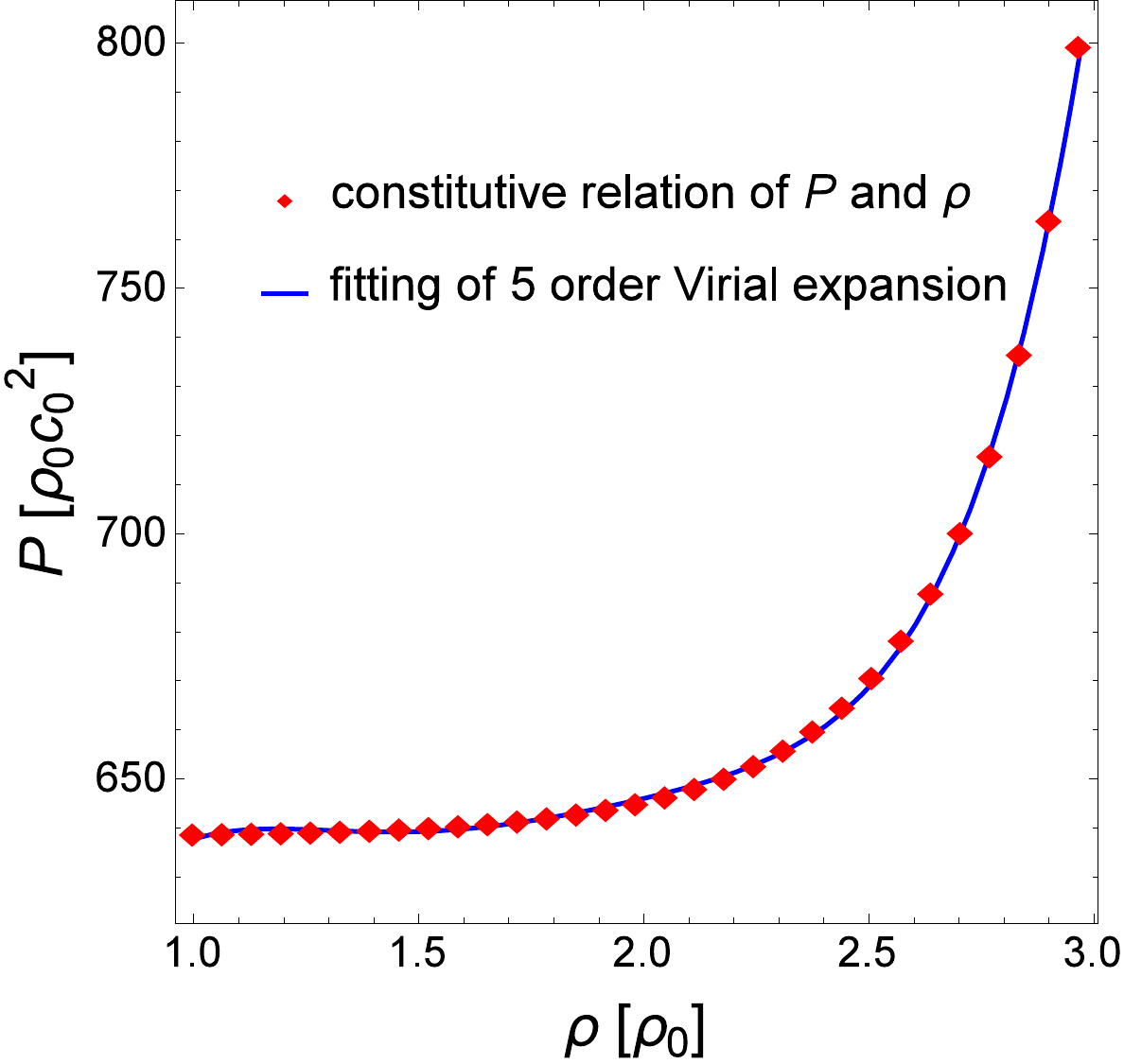}
\caption{\label{fig6}Constitutive relation between pressure $P$ and density $\rho$ fitted by 5 order Virial expansion (blue curve). The red dotted curve denotes the required constitutive relation corresponding to the unideal cloak of $\delta=0.1$ constructed by the transformation function Eq.~(\ref{f1}) with $n=4$.}
\end{figure}

Finally, we shall discuss the required state equation of the background fluid. Though the constitutive relation between $P$ and $\rho$ is determined by the coordinate transformation, we can use Virial expansion to approach our constitutive relation. The Virial expansion is a physical state equation that can be derived directly from statistical thermodynamics, which takes the form
\begin{equation}
  P=k_BT\rho+B_2(T)\rho^2+B_3(T)\rho^3+\cdots,
\end{equation}
where $k_B$, $T$ denote Boltzmann constant and temperature respectively, and $B_l$ is the $l$th order Virial coefficient. In Fig.~\ref{fig6}, we fit the $P-\rho$ relation for the unideal cloak of $\delta=0.1$, corresponding to Eq.~(\ref{f1}) with $n=4$, by 5 order Virial expansion. The fitting coefficients are $ B_1= 1994.73$, $B_2=-2430.51$, $B_3=1447.21$, $B_4=-422.527$, $B_5=48.7926$, and the pressure at infinity is $P_0=624.479$ (all coefficients are nondimensionalized in terms of the unit quantities $\rho_0$, $c_0$). Note that the expansion is not unique, since we can choose different orders to fit.

\section{Conclusion}

To summarize, we design an acoustic cloaking scheme by using background flow based on the method of ATA, and give all of the required physical parameters of the flow complying with the dynamic laws of fluid. We reveal that it is not enough to merely give the background velocity as previous works just did~\cite{garcia2013analogue,garcia2014space,garcia2014analogue}, but some restrictions should be satisfied, and some other necessities, such as the external mass and momentum sources, should be supplied. We also provide a comprehensive investigation about the propagation of sound waves in our cloaking system. According to our results, the major distinctive effect in our system is the phase retardation and lead when sounds bypass the cloak. Furthermore, we analyze the cloaking ability of unideal constructions and use Virial equation of state to fit the required constitutive relation of the background fluid. Our research not only offers a novel path to achieve acoustic cloaking but also complements the framework of ATA.





\end{CJK*}
\end{document}